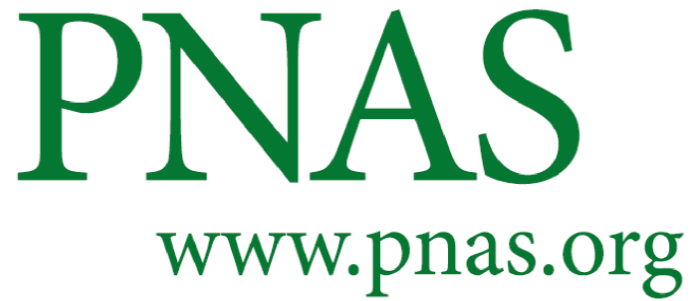

Main Manuscript for

**Linear regression does not encapsulate the effect of non-pharmaceutical interventions on the number of COVID-19 cases**

Angeline G. Pendergrass[1,2‡*], Kristie L. Ebi[3], and Micah B. Hahn[4]

[1] Institute for Atmospheric and Climate Science, ETH-Zurich, Zurich, Switzerland

[2] Climate and Global Dynamics Laboratory, National Center for Atmospheric Research, Boulder, CO, USA.

[3] Department of Global Health, University of Washington, Seattle, WA, USA

[4] Institute for Circumpolar Health Studies, University of Alaska-Anchorage, Anchorage, AK, USA

* Angeline G. Pendergrass

**Email:** apgrass@ucar.edu

AGP: https://orcid.org/0000-0003-2542-1461

KLE: https://orcid.org/0000-0003-4746-8236

MBH: https://orcid.org/0000-0002-7022-3918

**This PDF file includes:**

>Main Text
>Figure 1 and 2

---

[‡] This work was done during personal time as it is outside of her funded projects.

**Main Text**

Zhang et al. (1) argued that "mandated face covering represents the determinant in shaping the trends of the pandemic worldwide," and suggested that mandating masks in public is necessary and sufficient to decrease the rate of new COVID-19 infections. Their analysis assumed that the effect of the lockdowns in Italy and New York City (NYC) is captured by a linear increase in the number of cases (their Figs. 2b,c) that would have continued without mask mandates. We show two reasons why Zhang et al.'s statistical analysis of the association between mask mandates and SARS-CoV-2 transmission are inappropriate.

First, consider the susceptible-exposed-infected-recovered (SEIR) model (2), where $\beta$ is the rate at which contacts between the infected population fraction $i$ and susceptible fraction $s$ lead to infections (after a latent period), with subsequent recovery $r$. After introduction of coronavirus and before non-pharmaceutical interventions (NPIs), $\beta$ is high; NPIs, including both lockdowns and masks, act by decreasing $\beta$. The effect of one NPI, representing a lockdown alone, is shown in Fig. 1.

Zhang et al. quantified the effect of lockdowns as the slope of the cumulative number of cases, $N(i+r)$, where $N$ is constant population, fit with linear regression. Calculating $d(i+r)/dt$ from SEIR gives,

$$\frac{d}{dt}(i+r) = \beta s i.$$

$d(i+r)/dt$ is proportional not just to the effect of NPIs on $\beta$, but also to the number of interactions between susceptible and infected people. This changes over time, even when the effect of NPIs is constant (Fig. 1). Thus, the slope of the number of cases does not encapsulate the effect of NPIs, a central assumption of Zhang et al.'s analysis.

Second, consider a location which was not included in Zhang et al.'s analysis, Switzerland (Fig. 2). Switzerland implemented a lockdown broadly similar to those in Italy and NYC approximately two weeks after local transmission began (3, 4). However, unlike Italy and NYC, masks were not mandated for the general public (4) nor widely worn through June (5, 6). If mask mandates were the only factor that could drive a reduction in the number of cases, then the number of cases should have continued to steadily increase, as no mask mandate was in place. Instead, the rate of new cases peaked in late March and declined until early June (Fig. 2b), illustrating that NPIs other than mask mandates affect trends in COVID-19 cases.

Our approach shows that mask mandates may not be the only factor contributing to changes in COVID-19 trends, and other factors need to be considered when modelling changes in disease over time. It is thus likely that Zhang et al.'s analysis overestimated the number of new cases in Italy and NYC avoided by mask mandates alone. While masks appear to be an important element in reducing transmission of the virus (7), Zhang et al.'s analysis should not be considered sound evidence that mask mandates are sufficient to control or the primary factor controlling the spread of COVID-19.

**Acknowledgments**
Flavio Lehner, Dennis Hartmann, Reto Knutti, Paulo Ceppi, Brian Smoliak, Stephen Po-Chedley, and Céline Vetter provided feedback.

**Figure 1.** Timeseries of (a) total cases (c.f. Zhang et al. Figs. 1 and 2) and (b) new cases before and after an NPI simulated with an SEIR model. Parameters are latent period 4.55 days, infectious period 2.38 days, contact rate 1.58 day$^{-1}$ until day 45 and 0.35 day$^{-1}$ thereafter. Based on implementation by (8).

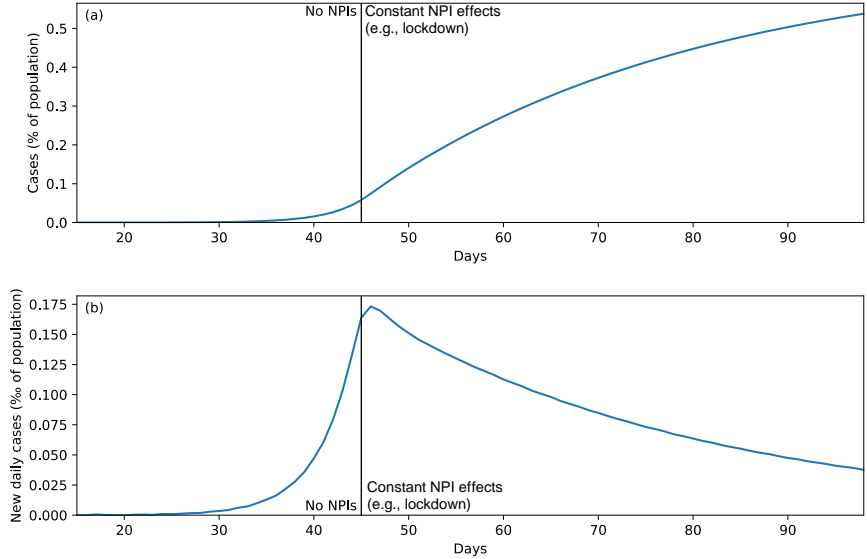

**Figure 2.** Timeseries of COVID-19 cases in Switzerland in spring of 2020. (a) Cumulative confirmed cases (c.f. (1) Fig. 2b,c). (b) Daily new confirmed cases (c.f. (1) Fig. 3). Raw data are shown in bars, the same data with the weekly cycle removed (8) and a 1-2-1 filter applied is shown by the line. Data from (9).

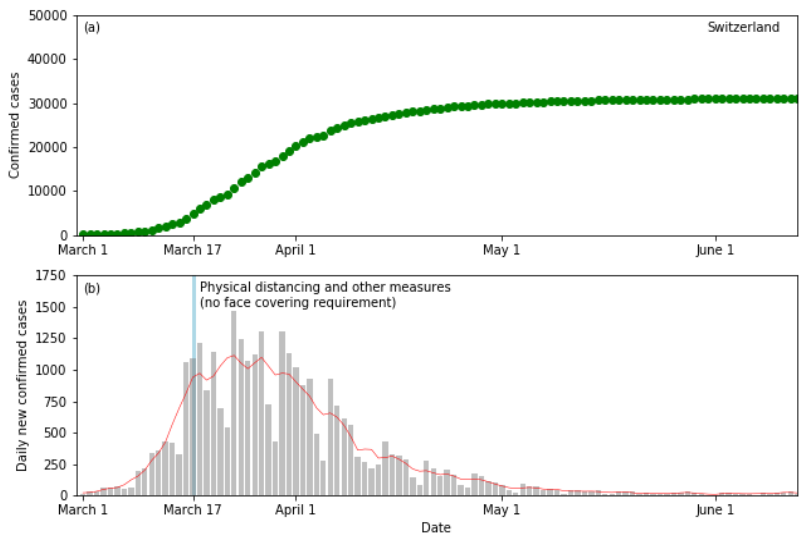